\documentclass[twocolumn,runningheads]{svjour2}
\smartqed  
\usepackage{graphicx}
\usepackage{epsfig}

\journalname{Astrophysics and Space Science}
\newcommand{\lnls}{log\,N($>$S)-log\,S}

\newcommand{\nh}{N$_{\rm H}$}

\newcommand{\rxdixhuit}{RX\,J1856.5\-$-$3754}
\newcommand{\rxzerosept}{RX\,J0720.4\-$-$3125}
\newcommand{\rxseize}{RX\,\-J16\-05.3\-+3249}

\newcommand{\rxa}{RX\,J0806.4\-$-$4123}
\newcommand{\rxb}{RX\,\-J0420.0\-$-$5022}
\def\aj{AJ}%
\def\apj{ApJ}%
\def\apjl{ApJ}%
\def\apjs{ApJS}%
\def\aap{A\&A}%
\def\mnras{MNRAS}%
\def\degr{\hbox{$^\circ$}}

\def\arcsec{\hbox{$^{\prime\prime}$}}
\def\pasp{PASP}%
\begin{document}

\title{Measuring proper motions of isolated neutron stars with Chandra.}
\subtitle{Constraints on the nature and origin of the ROSAT discovered isolated neutron stars}


\author{Christian Motch \and
        Adriana M. Pires \and
	Frank Haberl \and
	Axel Schwope
}


\institute{CNRS, UMR 7550, Observatoire Astronomique de Strasbourg, 11 rue de
           l'Universit\'e,  F-67000 Strasbourg, France
         \and
	 Instituto de Astronomia, Geof\'{\i}sica e Ci\^encias Atmosf\'ericas, Universidade de S\~ao Paulo, R. do Mat\~ao 1226, 05508-090 S\~ao Paulo, Brazil
	 \and 
	 Max-Plank Institut f\"ur Extraterrestrische Physik, Giessenbachstr., D-85748 Garching, Germany
	 \and
	 Astrophysikalisches Institut Potsdam, An der Sternwarte 16, D-14482 Potsdam, Germany
}

\date{Received: date / Accepted: date}

\maketitle

\begin{abstract}

The excellent spatial resolution of the Chandra observatory offers the unprecedented possibility to measure proper motions at X-ray wavelength with relatively high accuracy using as reference the background of extragalactic or remote galactic X-ray sources. We took advantage of this capability to constrain the proper motion of \rxa \ and \rxb, two X-ray bright and radio quiet isolated neutron stars (INSs) discovered by ROSAT and lacking an optical counterpart. In this paper, we present results from a preliminary analysis from which we derive 2$\sigma$ upper limits of 76 mas/yr and 138 mas/yr on the proper motions of \rxa\ and \rxb\ respectively. We use these values together with those of other ROSAT discovered INSs to constrain the origin, distance and evolutionary status of this particular group of objects. We find that the tangential velocities of radio quiet ROSAT neutron stars are probably consistent with those of 'normal' pulsars. Their distribution on the sky and, for those having accurate proper motion vectors, their possible birth places, all point to a local population, probably created in the part of the Gould Belt nearest to the earth.  

\end{abstract}

\section{Introduction}

The vast majority of isolated neutron stars (INS) are being discovered as radio pulsars. However, recent X-ray surveys have revealed that a number of these objects manifest themselves at X-ray and $\gamma$-ray wavelengths. A surprising result of the ROSAT all sky survey has been the discovery of a small group of INS exhibiting properties at variance from those of the more familiar radio-pulsars (see Haberl (2005) for a recent review). First, these neutron stars do not exhibit radio emission, although there are recent claims of the detection at a very low radio flux for two of these objects (Malofeev et al. 2006). Most of them display X-ray pulsations with periods close to 10\,s. Their X-ray spectra are essentially thermal with temperatures in the range of 40 to 100\,eV. Many ROSAT INSs display shallow broad absorption lines on the top of the thermal continuum, which are interpreted as proton cyclotron lines or as atomic transitions in high magnetic field conditions ( $\geq$ few 10$^{13}$ G). At present, it is still unclear whether or not ROSAT INSs constitute a homogeneous group, representative of a neutron star population different from standard radio-pulsars. The absence of strong radio emission can be due either to a very strong magnetic field (in excess of the quantum field) or to the fact that the radio pencil beam (which narrows at long spin periods) does not sweep the earth. Taking into account that only the closest sources have been detected by ROSAT because of their intrinsic X-ray faintness, their spatial density may well be comparable to that of normal radio-pulsars. ROSAT INSs may  be therefore just the tip of the iceberg, a formerly hidden large population of stellar remnants, possibly related to the growing number of radio-quiet and X-ray bright compact central objects in SNRs (see Pavlov et al. 2004 for a review). 

The generally low absorption (\nh $\sim 10^{20}\, {\rm cm}^{-2}$) derived from the soft X-ray observations suggest relatively small distances of the order of a few hundred parsecs. Small distances are indeed inferred from the HST parallax of the two X-ray brightest members, \rxdixhuit\ (d = 178$^{+22}_{-17}$\,pc, Kaplan 2003) and \rxzerosept\ (d = 330$^{+170}_{-80}$\,pc, van Kerkwijk \& Kaplan  2006). Proper motions in the range of 100 to 330\,mas\,yr$^{-1}$  are now measured for the three X-ray brightest ROSAT INSs (Walter 2001, Motch et al. 2003, Motch et al. 2005). These high velocities imply very low Bondi-Hoyle accretion rates and thus exclude with high confidence the possibility that the neutron star is re-heated by accretion of matter from the interstellar medium. The thermal emission of ROSAT INSs is most probably due to the progressive cooling of a middle-aged $\approx 10^5-10^6\, {\rm yr}$ neutron star. 

In this presentation we report on proper motion measurements of two ROSAT discovered INSs, \rxa\ and \rxb. These two neutron stars do not have established optical counterparts which could be used to measure their proper motions with optical telescopes. Instead we take advantage of the outstanding imaging quality of the Chandra observatory to constrain their proper motion using two observations  obtained in 2002 and 2005. The first part of the paper describes the observations and the method used to analyse them and to detect the displacement on the sky. In a second section, we use extensive MARX simulations to validate our results. We finally discuss the implication of our findings on the origin and distance of the group of thermally X-ray emitting but radio-quiet isolated neutron stars discovered by ROSAT.

\section{X-ray and optical properties of \rxa\ and \rxb} 

\rxa\ was discovered by Haberl et al. (1998) in the ROSAT all-sky survey with a PSPC count rate of 0.38 cts/s. Follow-up observations with the XMM-Newton satellite revealed pulsations at a period of 11.4\,s (Haberl \& Zavlin 2002, Haberl et al. 2004). Its thermal X-ray spectrum (kT = 96\,eV) lies among the hottest of this group. As for other ROSAT INSs, the quality of the spectral fit is significantly improved by adding a shallow absorption line centered around 460\,eV.  The source undergoes very little interstellar absorption (\nh\ = 4\,10$^{19}$\,cm$^{-2}$ or 1.1\,10$^{20}$\,cm$^{-2}$ depending on the inclusion or not of the low energy broad absorption line). Optical studies have been so far hampered by the low galactic latitude of \rxa \ ($b$ = -4.98\degr) and  no strong constraint exists on the optical brightness of the optical counterpart. ESO WFC imaging do not reveal optical counterparts brighter than an estimated R magnitude of 22 in the Chandra error circle. Deep radio observations failed to detect any source at a position consistent with that of \rxa \ (Johnston, 2003).

\rxb\ is a slightly fainter INS, also discovered in the ROSAT all-sky survey (Haberl et al.1999). Its PSPC count rate is 0.11 cts/s. XMM-Newton spectra show that the source is the coolest of its group with kT = 45\,eV and display evidences of a broad absorption line at E $\sim$ 330\,eV (Haberl et al. 2004). The interstellar absorption toward \rxb\  is \nh\ = 1.0\,10$^{20}$\,cm$^{-2}$ or 2.0\,10$^{20}$\,cm$^{-2}$, again depending on the inclusion or not of a broad line in the spectral model, and is on average about twice that of \rxa. Repeated XMM-Newton EPIC observations have now confirmed a pulsation period of 3.45\,s, the shortest of all ROSAT discovered INSs. A one hour long ESO-VLT exposure suggests the presence of a B = 26.6$\pm$0.3 mag object in the Chandra error circle, which owing to its faintness could well be the optical counterpart of the X-ray source. 

\section{Chandra Observations}

\subsection{Detecting proper motions}

We started in Cycle 3 a long term observing program aiming at measuring the motion of several ROSAT discovered INSs, either lacking optical counterparts, or with optical counterparts too faint to be measured from the ground in a reasonable amount of time.

A reference astrometric frame can be built from the background of AGNs with a relatively high accuracy. At low galactic latitudes, a fraction of the field sources are identified with remote galactic objects, mostly active stars, which typically have proper motions one or two orders of magnitude below that of the neutron star and can thus be used to build the reference frame. In order to minimize systematic effects, the 2002 and 2005 observations were acquired with the same instrumental setting (aim point, readout mode) and at similar times of the year and roll angles (see Table \ref{obs}). In both cases, the neutron star was located at the aim point of the instrument. The duration of the Chandra observation is determined by the need to acquire a large enough number of well positioned background extragalactic sources. In spite of the X-ray softness of \rxa, we preferred to use ACIS-I instead of a BI chip of ACIS-S to avoid strong pile up and eventual severe degradation of the positional accuracy of the INS as a result of the distortion of the PSF. For \rxb, which is significantly fainter and softer than \rxa, we chose to use ACIS-S to obtain the best position on the INS at the expense of a slightly smaller field of view. 

\begin{table}[t]
\caption{Journal of observations}
\centering
\label{obs}       
\begin{tabular}{lccc}
\hline\noalign{\smallskip}
Object & Observation & exposure & Roll angle\\[3pt]
       & date        & time (ks)& (deg)\\[3pt]
\tableheadseprule\noalign{\smallskip}
\rxa & 2002-02-23 & 17.7 & 322.4  \\
\rxa & 2005-02-18 & 19.7 & 325.6  \\
\rxb & 2002-11-11 & 19.4 & 18.6 \\
\rxb & 2005-11-07 & 19.7 & 23.6 \\
\noalign{\smallskip}\hline
\end{tabular}
\end{table}

\subsection{Source detection}

Data reduction and source detection was performed using CIAO 3.3.0.1 and the latest calibration database. The 2002 and 2005 observations were reprocessed in an homogeneous way and the best attitude corrections applied. The level-1 event files were corrected for known processing offsets and were reprocessed to the level-2 stage. The standard Chandra data reduction pipeline randomizes the X-ray event positions detected within a given pixel in order to remove the "gridded" appearance of the images and to avoid any aliasing effects. In principle, this randomization process could slightly degrade the source centering accuracy. We thus had to reprocess the raw data in order to remove the pixel randomization following the science thread available at the Chandra X-ray center. Source detection was run on both the normal standard processing data and de-randomized data. The {\em wavdetect} algorithm (wavelet transform) was used with pixel scales 1, 2 and 4 well suited to the detection of unresolved sources at moderate off-axis distances. We only considered sources detected in the central CCDs, chips 0 to 3 and 6 and 7 for ACIS-I and ACIS-S respectively. 

The final relative astrometric quality of the reference frames also depends on the number of sources common to the two observations and on the quality of the determination of their positions. We thus tried five different source detection thresholds, defined in {\em wavdetect} as the probability of a false detection at any given pixel. These thresholds ranged from 10$^{-8}$ up to a level of 5\,10$^{-5}$ in order to find the best compromise between the number of sources and the mean positional errors, both increasing with increasing threshold. The largest value implies that about 50 false sources per CCD chip enter the {\em wavdetect} source list. However, since we only consider sources detected in both the 2002 and 2005 observations and located at a maximum distance of one to three arcsec, the actual probability that the common source is spurious is very low. Best positions are obtained in the energy range in which the contrast between the source spectra and the unresolved extragalactic and instrumental backgrounds is optimal. For the neutron star, measured in the same conditions as the reference sources, the choice of the energy range is much less critical since the object is brighter than any of the background sources. We decided to test energy bands 0.3-10 keV, 0.3-5 keV, 0.3-3 keV, 0.5-5 keV and 0.5-2 keV. 

\subsection{Matching method}

Our relative astrometric frame determination uses the reconstructed equatorial positions of the X-ray sources. We allowed for translations of the right ascension and declination coordinates as well as for rotation around the aim point, resulting from possible errors in the attitude solution. A first algorithm simply uses a least square method ignoring the source positional errors. In a second step we implemented a maximum likelihood method consisting in maximizing the quantity;

$L\ = \ \sum_{i,j} \ \exp\,(-\frac{1}{2}\,(\frac{d_{ij}}{\sigma_{ij}})^{2})$ 

where $d_{ij}$ are the distances between sources $i$ and $j$ after transformation and $\sigma_{ij}$ the error on their distances, both quantities being computed independently for right ascension and declination. 

\subsection{Simulations}

To our knowledge, the Chandra data base does not contain repeated ACIS observations of high proper motion isolated neutron stars with suitable properties to be used as test data. In order to check our detection chain and constrain the resulting positional errors, we thus created several simulated data sets using the CXC ray-tracer \textsf{MARX 4.2.1} . 

A total of 26 and 12 sources were simulated for the fields of \rxa\ and \rxb , respectively. These sources are those common (within 3\arcsec) to the actual 2002 and 2005 observations as detected in the 0.5-5.0\,keV band and with a threshold of 10$^{-6}$ for \rxa\ and in the 0.5-2.0\,keV band with a threshold of 10$^{-5}$ for \rxb. We assumed a power law spectral energy distribution, typical of that of the extragalactic population which is known to be dominant on the X-ray background. Theoretical spectra of an absorbed power law of spectral index $\Gamma = 1.7$, undergoing a hydrogen column density of \nh$ = 5.21 \, 10^{21}$ cm$^{-2}$ (for the field of \rxa) and \nh$ = 1.07 \, 10^{20}$ cm$^{-2}$ (for the field of \rxb), were created using \textsf{XSPEC 12.2.1}. The column density values were derived from the colour excess in the direction of the sources as provided by the maps of Schlegel et al. (1998) available at the NASA/IPAC Extragalactic Database -- respectively, $\rm E_{\rm (B-V)} = 0.971$ and $\rm E_{\rm (B-V)} = 0.020$ -- and applying the Predehl \& Schmitt (1995) relation between the optical extinction and the X-ray absorption, \nh$ = (1.79 \pm 0.03) \rm A_V\,10^{21}$ cm$^{-2}$. On the other hand, the neutron stars were assumed to radiate as blackbodies with $kT = 96$\,eV and $kT = 44$\,eV (Haberl 2005). 

Each source was individually simulated giving as input parameters its equatorial coordinates, photon flux and spectral energy distribution, as well as the observation exposure and start time (in order to apply the proper degree of ACIS hydrocarbon contamination that is continuously degrading the quantum efficiency of the CCDs, specially at energies below 1\,keV), the detector type at the focal plane and the coordinates of the aim-point. The flux normalization of each source was computed converting the ACIS count rates into flux in units of photons\,s$^{-1}$\,cm$^{-2}$ in the 0.1-10.0\,keV energy band. Events generated for each source were then merged into a single event file, which mimics the spatial and flux distribution of the actual sources present in the fields of \rxa \ and \rxb. Background X-ray events were extracted from experimental blank fields provided by the calibration database. These blank fields were reprojected onto the tangent plane of the observations and, for each simulation, the events were randomly picked from the map so as to reproduce the level of background noise of the observation in the 0.5-5.0 keV band. Finally, sources and background were merged into a single event file. 

The count distribution of the artificial sources is in good agreement with a Poisson distribution. We had to adjust the input flux of each source simulation to correct for the exposure map, vignetting and discrepant counts for the few sources located close to CCD edges or bad pixels and columns. It also seems that the {\em wavdetect} algorithm systematically underestimates by a few percents or more the number of X-ray photons and the input source parameters in \textsf{MARX} had to be increased accordingly to recover a distribution in source counts similar to that of the actual observations. 

A total of 14 pairs of simulated event lists were generated for each ACIS I and S detector configurations. The main parameters changed were i) the displacement applied to the central source, 0, 0.34, 0.75 and 1\arcsec, ii) the offset applied to the 2005 reference coordinate system with respect to the 2002 one, 0 and 0.29\arcsec\ and iii) the position of the aim point which was moved by the equivalent of 1/4, 1/2 and 2/3 of a pixel in order to produce different distributions of the source PSF on the CCD pixel grid, although the Lissajous shaped dithering should in principle wash out a large part of the effect. In each of the 14 configurations, both event lists with and without pixel randomization applied were considered.

We discuss here preliminary results of the ACIS-I simulations corresponding to the \rxa \ observations. A detailed analysis of all ACIS-I and S simulations will be presented in a forthcoming paper.

Owing to its large number of photons, the formal errors on the position of the isolated neutron star delivered by {\em wavdetect} are very small, of the order of 0.02\arcsec. In actual data, however, one should also take into account a small additional error which results from uncertainties in the transformation between the detector pixel and sky coordinates. Deep ACIS-I observation of a field in Orion containing over a thousand of bright X-ray sources show that this additional error is of the order of $\sim$ 0.07\arcsec\ (Chandra Proposer's Observatory Guide). Therefore, most of the uncertainty on the proper motion probably comes from the accuracy with which we can link the reference frames of the two epochs. This accuracy is given by $\sigma_{\rm frame}^{2} = \frac{1}{n (n-1)} \sum_{i=1}^{n} (d_{i}^{12})^{2}$ with $d_{i}^{12}$ being the distance between source $i$ as observed in the first and second observation. Most of the configurations yield a $\sigma_{\rm frame}$ close to 0.10\arcsec. Best results are obtained when restricting the energy band in which source detection is done to 0.3-5.0 keV or 0.5-5.0 keV. 

\begin{figure}

\psfig{figure=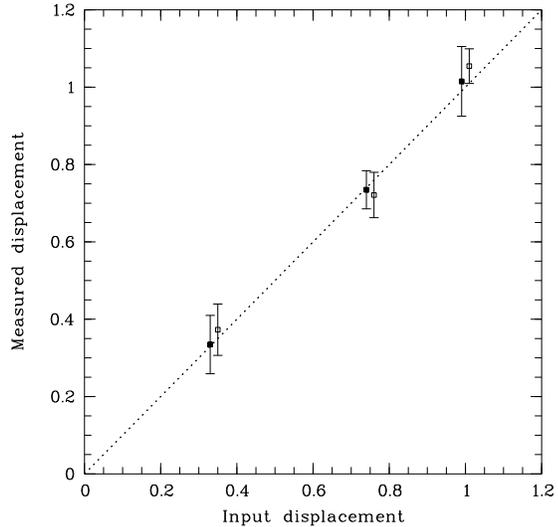,width=8cm,angle=-90,bbllx=5pt,bblly=74pt,bburx=600pt,bbury=620pt,clip=true}

\caption[]{Measured displacement as a function of the displacement entered in the simulations. Energy band 0.3-5 keV and 0.5-5 keV. Filled squares and open squares represent simulations with and without pixel randomization (small offsets applied to the input values for clarity).}

\label{sim1}
\end{figure}

In our simulations, the pixel randomization process has basically no influence on the accuracy of the frame matching and there is no strong dependency either on the threshold used for source detection. We show in Fig. \ref{sim1} the relation between the input displacements and their values measured on the simulated data. On the average, the motion is recovered with an rms accuracy of the order of 0.07 to 0.10\arcsec. In a similar manner, the shifts in right ascension and declination are recovered with a rms accuracy of 0.07\arcsec. 

\subsection{Results}

As for the \textsf{MARX 4.2.1} simulations, the $\sigma_{\rm frame}$ for real data are minimum for the energy ranges 0.3-5.0 keV and 0.5-5.0 keV for ACIS-I and 0.3-3.0 and 0.5-5.0 keV for ACIS-S. The number of sources common to both observations varies with detection threshold, energy range and maximum allowed distance between the 2002 and 2005 positions to include the source in the computation of the transformation. It ranges from 12 to 39 sources for \rxa \ and from  6 to 16 sources for \rxb. Introducing sources more distant than 1\arcsec\ (at large off-axis angles) somewhat worsens the quality of the relative astrometry. Detection thresholds smaller than 10$^{-6}$ give results of lower quality as a consequence of the decreasing number of real sources detected at the two epochs. Using the best combination of energy bands and detection thresholds, the mean distance between reference sources after transformation is 0.46\arcsec\ for ACIS-I and 0.57\arcsec\ for ACIS-S, again with no strong dependency on the fact that pixel randomization is applied or not. These mean distances are slightly larger by $\sim$ 30\% than those expected from the formal errors given by {\em wavdetect}. A small offset of 0.14\arcsec\ in right ascension and 0.39\arcsec\ in declination is found between the 2002 and 2005 coordinate systems of the field of \rxa, while no really significant translation or rotation is needed between the equatorial systems of the two observations of \rxb.

For \rxa, the measured displacement averaged over the best energy bands and detection thresholds is $\sim$ 0.1\arcsec, corresponding to a $\sim$ 1$\sigma$ effect. The error on the transformation, $\sigma_{\rm frame}$ probably dominates the error budget. With a best value of  $\sigma_{\rm frame}$ = 0.09\arcsec\ and assuming an additional systematic error of 0.07\arcsec, the 3$\sigma$ upper limit on the total displacement of \rxa \ over the 3 years time interval is 0.34\arcsec\ implying $\mu\ \leq$ 114 mas/yr (or $\mu\ \leq$ 76 mas/yr at the 2$\sigma$ level).

Because of the smaller central field of ACIS-S and accordingly smaller number of reference sources, the mean $\sigma_{\rm frame}$ is 0.19\arcsec\ for the field of \rxb.  The mean displacement measured is 0.09\arcsec\ corresponding to a $\sim$ 0.4$\sigma$ effect. For \rxb\ too, we do not measure any statistically significant motion and can put a 3$\sigma$ upper limit of $\mu$ = 207 mas/yr (or $\mu\ \leq$ 138 mas/yr at the 2$\sigma$ level). 


Our observing programme has thus established the possibility to measure proper motions at X-ray wavelengths with relatively good accuracy. In particular, the proper motions of \rxdixhuit, \rxseize\ and possibly \rxzerosept\ would have been detected with ACIS-I and a three year time interval. If Chandra lives long enough, very significant constraints or detections of the proper motions of the ROSAT discovered INSs that are too faint to be observed with optical telescopes will be possible. 

\section{Discussion}

The upper limit on the proper motions of \rxa \ and \rxb\ derived from our Chandra observations provide significant constraints on the space velocities of these two objects. In particular, the low value of 76 mas/yr obtained for \rxa\ is well below those measured so far for the three X-ray brightest objects but comparable to those of nearby radio pulsars (see Table \ref{propermotions}). The likely small distance of the source ($d$ $\leq$ 240\,pc, Posselt et al. 2006) imply low transverse velocities of less than 83 km\,s$^{-1}$. 

\begin{center}
\begin{table}
\caption[]{Proper motions, distances and transverse velocities of nearby young neutron stars}
\label{propermotions}
\begin{tabular}{llll}
\noalign{\smallskip}
Name            &   Proper motion  &  distance  	& V$_{\rm T}$  \\ 
                &   (mas/yr) 	   &	(pc)		& (km\,s$^{-1}$) \\
\hline
\rxdixhuit      &   332$\pm$1	   & 178$^{+22}_{-17}$  & 283 \\ 
\rxzerosept     &   97$\pm$12	   &  250		& 114 \\ 
\rxseize        & 144.5$\pm$13.2   &  $<$ 410		& $<$ 280\\ 
\rxa            & $\leq$ 76        & 240$\pm$25         & $<$ 83\\
\rxb            & $\leq$138        & $\leq$340          & $<$ 241\\
\hline
B0656+14        &  44.1$\pm$0.7    & 288$^{+33}_{-27}$  & 60.1 \\ 
B1929+10        & 103.4$\pm$0.2    &361$^{+10}_{-8}$	& 176 \\ 
Vela Pulsar     & 58.0$\pm$0.1	   & 287$^{+19}_{-17}$  & 79\\ 
Geminga         & 170$\pm$4 	   & 273$\pm$84 	& 219\\ 
\noalign{\smallskip}
\hline
\end{tabular}
\end{table}
\end{center}

\begin{figure}

\psfig{figure=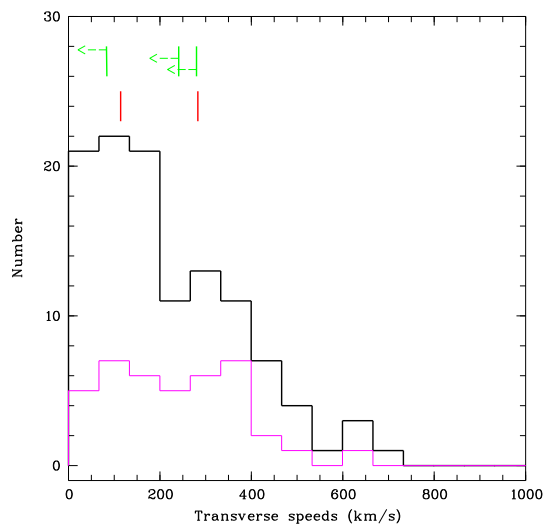,width=8cm,angle=-90,bbllx=5pt,bblly=74pt,bburx=600pt,bbury=620pt,clip=true}

\caption[]{Transverse speeds of ROSAT discovered INSs compared to those of non recycled radio pulsars. Data taken from Hobbs et al. (2005). The black thick line and the thin magenta line show the histogrammes of the entire population and of the young (age $\leq$ 3 10$^{6}$yr) respectively. Measured values and upper limits for the ROSAT INSs are shown on the top.}

\label{fig_histogramme_transverse_speeds-hobbspaper_lin}
\end{figure}

On Fig. \ref{fig_histogramme_transverse_speeds-hobbspaper_lin} we compare the transverse velocities of ROSAT discovered INSs with those of non recycled radio pulsars, total and young (age $\leq$ 3 10$^{6}$yr) populations. Radio pulsar data are taken from Hobbs et al. (2005). The distribution of the inferred 2-D velocities of the ROSAT INSs appears consistent with that of the radio pulsars. The absence of fast moving ROSAT INSs is intriguing but not really statistically significant, since the probability of such a combination remains above the 10\% level. According to Hobbs et al. (2005), young radio pulsars have a mean 2-D velocity slightly larger than old ones. However, the two distributions overlap, as seen on Fig. \ref{fig_histogramme_transverse_speeds-hobbspaper_lin}, and the significance of this difference remains uncertain. The ATNF catalogue (Manchester et al. 2005) contains a total of 6 radio pulsars with $d$ $\leq$ 1\,kpc, younger than 4.25 Myr and with measured proper motion. They have 2-D velocity in the range of 60 to 354 km\,s$^{-1}$ with a mean of 177 km\,s$^{-1}$, clearly consistent with that observed for the ROSAT INSs.

The measurement of proper motions can potentially provide a wealth of important informations on the evolutionary status, origin and age of the neutron star. A large spatial velocity, typically in excess of 30 km\,s$^ {-1}$ rules out the possibility that the star is reheated by accretion of interstellar material. 

In this respect, the upper limits on the 2-D velocities of \rxa\ and \rxb\ are still large and the possibility that they accrete from the interstellar medium cannot yet be seriously considered. The fact that these two INSs share many properties such as X-ray energy distribution and pulsation periods with high spatial velocity ROSAT INSs also argues for a similar nature.  

If measured with enough accuracy, the proper motion vector of the neutron star can also be used to compute possible past trajectories assuming a range of present distances and radial velocities. OB stars progenitors of neutron stars are not distributed randomly on the sky but rather concentrate in relatively large associations sharing a common space motion. The exact fraction of early type stars found in OB associations remains in many cases uncertain due to the errors on the photometric distances and the generally fuzzy boundaries attributed to the associations. Garmany (1994) shows that the relative fraction of O and B stars found in associations is larger than 75\% and 58\% respectively. The fact that a group of backwards trajectories crosses the past position of a known OB association is thus a hint, no definite proof since other trajectories are possible, that the neutron star progenitor was located there when it underwent supernova explosion. It becomes then possible to compute possible travel times which can be compared to theoretical cooling curves, with the spin down time and, if associated to a SNR, with the expansion time. For seven nearby and young isolated neutron stars listed in Table \ref{birthplace} a tentative birth place has been proposed in the recent literature. 

\begin{figure*}

\psfig{figure=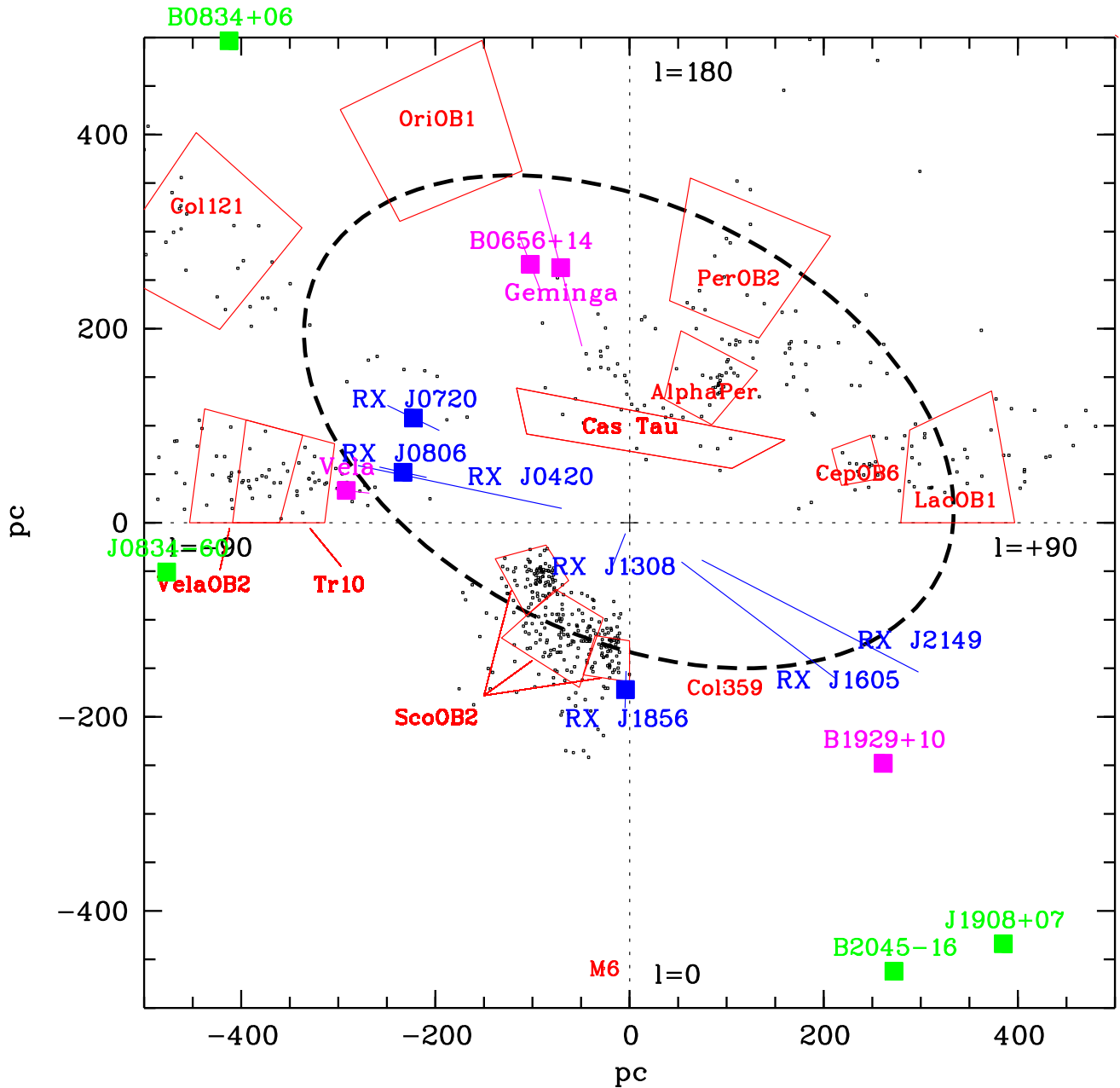,height=18cm,angle=-90,bbllx=5pt,bblly=74pt,bburx=600pt,bbury=620pt,clip=true}

\caption[]{Positions of nearby neutron stars and OB associations projected on the Galactic plane. Red boxes show the OB associations boundaries. Hipparcos stars with a probability higher than 75\% to be linked to the OB associations are represented by black dots. Blue lines and filled squares show the possible positions of the ROSAT discovered INS, assuming a distance range of 100 to 400\,pc for those which do not have distance estimates. Radio or $\gamma$-ray pulsars younger than 4.25 Myr and located within 1\,kpc are shown as magenta symbols (when a parallax distance exists) or green symbols when the distance is estimated from dispersion measurements. The Gould Belt is shown as a thick dashed line.}

\label{plotOBMembersAndINSinPlane}
\end{figure*}

We show on Fig. \ref{plotOBMembersAndINSinPlane} the positions projected on the galactic plane of neutron stars younger than 4.25\,Myr and located at distances less than 1\,kpc. Radio pulsar data were extracted from the pulsar catalogue maintained by the Australian Telescope National Facility (Ma\-nchester et al. 2005). We also show the boundaries of the classical OB associations as defined in Humphreys (1978) and de Zeeuw et al. (1999). For the Gould Belt, we use the determination of Perrot \& Grenier (2003) based on HI and H$_{2}$ clouds and Hipparcos distances to the nearby OB associations.

Neutron star travel times from their supposed birth place to their present location are difficult to determine with great accuracy. The large extents of the possible parent association and the unknown radial velocities usually yield rather large uncertainties.  In the case of the two youngest nearby radio pulsars, the Vela pulsar (10$^{4}$yr) and B0656+14 (10$^{5}$yr), there is a good agreement between the estimated travel time, the pulsar age computed assuming magnetic dipole braking and the age of the remnant of the parent supernova (Hoogerwerf et al. 2001; Thorsett et al. 2003). The absence of obvious supernova remnants at the possible birth place of  B1929+10 and Geminga is consistent with their older ages. In this respect, the lack of conspicuous remnants in the surroundings of the hypothetical birth locations of the ROSAT discovered INSs is also consistent with the relatively large ages ($\geq$ 3\,10$^{5}$yr) derived from the flight times (see Table \ref{birthplace}).     

\begin{center}
\begin{table*}[ht]
\caption[]{Possible birth places of nearby young neutron stars}
\label{birthplace}
\begin{tabular}{lllll}
\noalign{\smallskip}
Name           & spin down Age  & Travel time	   &  Birth place & Reference\\
               & (yr)		& (yr)  	   &		 &  \\
\hline
\rxdixhuit     &  ?		& 4\,10$^{5}$ - 1\,10$^{6}$    & Upper Sco OB2     &  1,2\\
\rxzerosept    &  2\,10$^{6}$   &  6\,10$^{5}$ - 3\,10$^{6}$    & Tr 10 + Vela OB2 or Lower Sco OB2 & 3,11 \\
\rxseize       &  ?		& $\sim$10$^{6}$ &   upper Sco OB2 &  4 \\
\hline
B0656+14       &  10$^{5}$	&  10$^{5}$	 &  Monogem Ring   &  5,6 \\
B1929+10       &  3\,10$^{6}$   &  1\,10$^{6}$  & Upper Sco OB2	& 7,8\\
Vela Pulsar    &  10$^{4}$	&  10$^{4}$	 &  Vela OB2  &  9,8 \\
Geminga        &  3.4\,10$^{5}$ &  3.4\,10$^{5}$ & Cas-Tau or Ori OB1 &  10\\
\noalign{\smallskip}
\hline
\end{tabular}

1) Kaplan 2003; 
2) Walter 2001; 3) Motch et al. 2003; 
4) Motch et al. 2005; 5) Brisken et al. 2003; 6) Thorsett et al. 2003; 7) Chatterjee et al. 2004; 8)Hoogerwerf et al. 2001; 9) Dodson et al. 2003; 10) Pellizza et al. 2005; 11) Kaplan \& van Kerkwijk 2005.
\end{table*}
\end{center}

According to Perrot \& Grenier (2003), the Gould Belt has an elliptical shape with major and minor semi-axes of 373\,$\pm$\,5 pc and 233\,$\pm$\,5\,pc respectively. The Sun lies inside the Belt and is located about 104\,$\pm$\,4\,pc offset from its center. It is currently closer to the region of the Gould Belt populated by the Sco OB2 and Vela OB2 + Trumpler 10 associations (see Fig. \ref{plotOBMembersAndINSinPlane}). Backwards trajectories suggest an origin in the closest region of the Belt for \rxdixhuit, \rxzerosept, \rxseize\ and B1929+10. Two INSs, Geminga and B0656+14 were possibly born in or close to the Orion complex, on the opposite side of the Gould Belt. As visible in Fig. \ref{plotOBMembersAndINSinPlane}, all ROSAT discovered INSs lie in the half sky centered on Sco OB2 and containing the closest part of the Gould Belt. 

No such sensitivity anisotropy exists in the ROSAT all sky survey at the level of 0.14 cts/s, the PSPC count rate of \rxb \ which is the faintest of the ROSAT discovered INSs (Voges et al. 1999). In order to detect and recognise the extreme softness of such a source a minimum of $\sim$ 200\,s exposure time is needed. Only 10\% of the whole sky had exposures shorter than this value. Furthermore, the distribution of the low exposed regions are not specifically directed towards the far side of the Gould Belt where ROSAT INSs are apparently not detected. 

The very soft thermal-like energy distributions of the ROSAT INSs render their detectability very sensitive to the amount of interstellar absorption on the line of sight. The distribution of the local interstellar medium has been the subject of extensive studies, mostly based on the measurement of the NaID doublet at 5890\AA\ (Welsh et al. 1994; Sfeir et al. 1999). Our current environment is rather unusual since we appear to be located in a local bubble cavity of very low interstellar gas density. The bubble is surrounded by a "wall" of interstellar matter. Its shape is very irregular with the closest distance to the edge of the bubble being at $\sim$ 60\,pc in the direction of the Galactic center. Being based on more than a thousand of directions, the NaI absorption map of Lallement et al. (2003) provides a good description of the local interstellar medium up to distances of about 350\,pc for absorptions up to a few 10$^{20}$ cm$^{-2}$. Their map shows no evidence for any asymmetry which could explain the absence of detection of INSs in the half sky where they apparently miss. 

Therefore, although based on small number statistics, the distribution of ROSAT discovered INSs on the sky might well reveal the major role played by the local Gould Belt in their generation. Popov et al. (2003) have indeed highlighted the importance of the Gould Belt for understanding the bright part of the observed X-ray \lnls\ INS curve.

\section{Conclusions}

The high astrometric quality of the Chandra Observatory offers the possibility to measure proper motions of X-ray sources with an unprecedented accuracy. Our observations and simulations show that displacements as small as $\sim$ 0.3\arcsec\ can be detected. Applying this method to two ROSAT discovered INSs, we derive an upper limit on the transverse speed of \rxa\ at the lower end of the observed distribution for radio pulsars while less constraining limits are obtained for \rxb. We show that the spatial distribution and possible birth places of this particular group of INSs argue in favour of a local population and for a production dominated by the closest part of the Gould Belt.

\begin{acknowledgements}
We acknowledge the use of the ATNF Pulsar Catalogue available at http://www.atnf.csiro.au/ \-re\-search/pulsar/psrcat. 
\end{acknowledgements}


\begin{thebibliography}{}

\bibitem{2003ApJ...593L..89B} Brisken, W.~F., 
Thorsett, S.~E., Golden, A., \& Goss, W.~M.\ 2003, \apjl, 593, L89 

\bibitem{2004ApJ...604..339C} Chatterjee, S., 
Cordes, J.~M., Vlemmings, W.~H.~T., Arzoumanian, Z., Goss, W.~M., \& Lazio, 
T.~J.~W.\ 2004, \apj, 604, 339 

\bibitem{1999AJ....117..354D} de Zeeuw, P.~T., 
Hoogerwerf, R., de Bruijne, J.~H.~J., Brown, A.~G.~A., \& Blaauw, A.\ 1999, 
\aj, 117, 354 

\bibitem{2003ApJ...596.1137D} Dodson, R., Legge, D., 
Reynolds, J.~E., \& McCulloch, P.~M.\ 2003, \apj, 596, 1137 

\bibitem{1994PASP..106...25G} Garmany, C.~D.\ 1994, \pasp, 106, 25 

\bibitem{1998AN....319...97H} Haberl, F., Motch, C., 
\& Pietsch, W.\ 1998, Astronomische Nachrichten, 319, 97 

\bibitem{1999A&A...351L..53H} Haberl, F., Pietsch, W., 
\& Motch, C.\ 1999, \aap, 351, L53 

\bibitem{2002A&A...391..571H} Haberl, F., \& 
Zavlin, V.~E.\ 2002, \aap, 391, 571 

\bibitem{2004A&A...424..635H} Haberl, F., Motch, C., Zavlin, V.E. et al.\ 
2004, \aap, 424, 635 


\bibitem{2005fysx.conf...39H} Haberl, F.\ 2005, 5 years of 
Science with XMM-Newton, MPE Report 288, 39 


\bibitem{2005MNRAS.360..974H} Hobbs, G., Lorimer, 
D.~R., Lyne, A.~G., \& Kramer, M.\ 2005, \mnras, 360, 974 


\bibitem{2001A&A...365...49H} Hoogerwerf, R., de 
Bruijne, J.~H.~J., \& de Zeeuw, P.~T.\ 2001, \aap, 365, 49 

\bibitem{1978ApJS...38..309H} Humphreys, R.~M.\ 1978, 
\apjs, 38, 309 

\bibitem{2003MNRAS.340L..43J} Johnston, S.\ 2003, \mnras, 
340, L43 

\bibitem{}Kaplan, D., 2003 in proceedings of the workshop "Physics and Astrophysics of Neutron Stars, Jult 28- August 1 2003, Santa Fe, New Mexico.

\bibitem{2005ApJ...628L..45K} Kaplan, D.~L., \& van Kerkwijk, M.~H.\ 2005, \apjl, 628, L45
 

\bibitem{2003A&A...411..447L} Lallement, R., Welsh, 
B.~Y., Vergely, J.~L., Crifo, F., \& Sfeir, D.\ 2003, \aap, 411, 447 

\bibitem{}Malofeev, V., Malov, O., Teplykh, D., 2006, this conference

\bibitem{}Manchester, R. N., Hobbs, G. B., Teoh, A. \& Hobbs, M., Astron. J., 129, 1993-2006 (2005)

\bibitem{2003A&A...408..323M} Motch, C., Zavlin, V.~E., \& Haberl, F.\ 2003, \aap, 408, 323 

\bibitem{2005A&A...429..257M} Motch, C., Sekiguchi, K., 
Haberl, F., Zavlin, V.~E., Schwope, A., \& Pakull, M.~W.\ 2005, \aap, 429, 
257 

\bibitem{2004IAUS..218..239P} Pavlov, G.~G., Sanwal, 
D., \& Teter, M.~A.\ 2004, IAU Symposium, 218, 239 


\bibitem{2005A&A...435..625P} Pellizza, L.~J., 
Mignani, R.~P., Grenier, I.~A., \& Mirabel, I.~F.\ 2005, \aap, 435, 625 

\bibitem{2003A&A...404..519P} Perrot, C.~A., \& 
Grenier, I.~A.\ 2003, \aap, 404, 519 

\bibitem{} Popov, S.B., Colpi, M., Prokhorov, M.E., Treves, A., Turolla, R., 2003, \aap, 406, 111

\bibitem{}Posselt et al. 2006, this conference


\bibitem{1995A&A...293..889P} Predehl, P., \& 
Schmitt, J.~H.~M.~M.\ 1995, \aap, 293, 889 

\bibitem{1998ApJ...500..525S} Schlegel, D.~J., 
Finkbeiner, D.~P., \& Davis, M.\ 1998, \apj, 500, 525 

\bibitem{1999A&A...346..785S} Sfeir, D.~M., Lallement, 
R., Crifo, F., \& Welsh, B.~Y.\ 1999, \aap, 346, 78

\bibitem{2003ApJ...592L..71T} Thorsett, S.~E., 
Benjamin, R.~A., Brisken, W.~F., Golden, A., \& Goss, W.~M.\ 2003, \apjl, 
592, L71 

\bibitem{}van Kerkwijk, M. H.\& D. L. Kaplan, 2006, this conférence

\bibitem{1999A&A...349..389V} Voges, W., et al.\ 1999, \aap, 349, 389
 
\bibitem{2001ApJ...549..433W} Walter, F.~M.\ 2001, \apj, 549, 
433 


\bibitem{1994ApJ...437..638W} Welsh, B.~Y., Craig, N., 
Vedder, P.~W., \& Vallerga, J.~V.\ 1994, \apj, 437, 638 
 
\end{thebibliography}
\end{document}